\documentclass[
    ,final            
  ]
  {aipproc}

\layoutstyle{6x9}

\usepackage{amssymb, amsmath}

\newcommand{\bi}{\begin{itemize}}
\newcommand{\ei}{\end{itemize}}

\newcommand{\be}{\begin{equation}}
\newcommand{\ee}{\end{equation}}
\newcommand{\bea}{\begin{eqnarray}}
\newcommand{\eea}{\end{eqnarray}}

\begin{document}

\title{Hadronic matrix elements for $B$-mixing in the Standard Model and beyond}


\author{C.~M.~Bouchard,\,  {\it for the Fermilab Lattice and MILC Collaborations}}{
  address={Physics Department, The Ohio State University, Columbus, Ohio 43210, USA}
}

\begin{abstract}
We use lattice QCD to calculate the $B$-mixing hadronic matrix elements for a basis of effective four-quark operators that spans the space of all possible contributions in, and beyond, the Standard Model.  We present results for the SU(3)-breaking ratio $\xi$ and discuss our ongoing calculation of the mixing matrix elements, including the first calculation of the beyond the Standard Model matrix elements from unquenched lattice QCD.
\end{abstract}

\maketitle


\section{Motivation}
Through a combination of the GIM mechanism, Cabibbo suppression, and loop suppression, the Standard Model (SM) contribution to $B$-mixing is small and new physics (NP) effects may be discernible~\cite{Buras:2012}.  
In fact, there are experimental hints this may be the case.  In unitarity triangle analyses \cite{Laiho:2011b,Lunghi:2011} a persistent $2-3\sigma$ inconsistency is suggestive of NP and points to $B$-mixing as a possible source.  D$\emptyset$'s same-sign dimuon charge asymmetry~\cite{Abazov:2011} and global analyses by UTfit~\cite{Bevan:2010} and CKMfitter~\cite{Lenz:2012} reveal $2-4 \sigma$ discrepancies in SM $B$-mixing.
Experimental measurements of the $B$-mixing oscillation frequency~\cite{HFAG:2012} have sub-percent precision but cannot be fully leveraged in the search for NP as theory errors, dominated by hadronic uncertainty, are an order of magnitude larger~\cite{Lenz:2011}.

\section{Calculation}
\begin{figure}[t!]
  \includegraphics[height=.13\textheight]{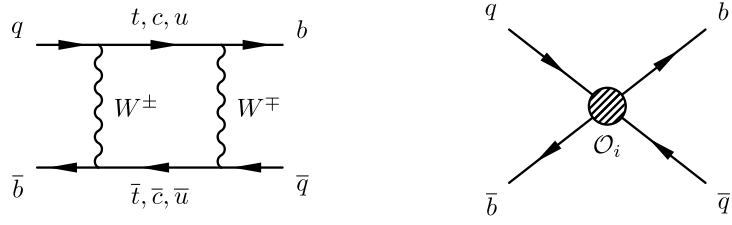}
  \caption{({\it Left}) A SM contribution to $B$-mixing and ({\it right}) an effective four-quark interaction.}
  \label{fig:feyndiag}
\end{figure}

To lowest order in the SM, $B$-mixing is described by box diagrams [{\it cf.} Fig.~\ref{fig:feyndiag} ({\it left})].  Under the operator product expansion (OPE) flavor-changing short-distance interactions, of O(100 GeV) in the SM and higher in NP scenarios, and long-distance hadronic physics, of O(500 MeV), factorize.  At the energies relevant to $B$-mixing, of O($M_B\sim 5\ $GeV), flavor-changing physics is described by the local, effective, four-quark interaction of Fig.~\ref{fig:feyndiag} ({\it right}).
A commonly used basis of mixing operators is
\begin{eqnarray}
{\cal O}_1 &=& (\bar{b}^\alpha \gamma_\mu L q^\alpha) \ (\bar{b}^\beta \gamma_\mu L q^\beta) \hspace{0.5in} {\cal O}_4\ \ =\ \ (\bar{b}^\alpha L q^\alpha) \ (\bar{b}^\beta R q^\beta) \nonumber \\
{\cal O}_2 &=& (\bar{b}^\alpha L q^\alpha) \ (\bar{b}^\beta L q^\beta) \hspace{0.82in} {\cal O}_5\ \ =\ \ (\bar{b}^\alpha L q^\beta) \ (\bar{b}^\beta R q^\alpha) \nonumber \\
{\cal O}_3 &=& (\bar{b}^\alpha L q^\beta) \ (\bar{b}^\beta L q^\alpha)
\label{mixops}
\end{eqnarray}
where $L/R$ are left/right projection operators and $\alpha, \beta$ are color indices.  A calculation of the matrix elements of these operators is sufficient to parameterize the hadronic contributions to $B$-mixing in and beyond the SM.

Under the OPE, the expression for the oscillation frequency factorizes:
\begin{equation}
\Delta M_q = \sum_i C_i(\mu)\ \langle B_q^0|{\cal O}_i(\mu)|\bar{B}^0_q\rangle,
\label{dM_BSM}
\end{equation}
where the short-distance $C_i$ are model dependent and the long-distance hadronic mixing matrix elements $\langle B_q^0|{\cal O}_i|\bar{B}^0_q\rangle$ must be calculated nonperturbatively using lattice QCD.  The phenomenologically-relevant bag parameters $B^{(i)}_{B_q}$ and SU(3)-breaking ratio $\xi$ are related to the matrix elements by
\begin{equation}
 \langle B_q^0|{\cal O}_i|\bar{B}^0_q\rangle = {\mathfrak c}_i f_{B_q}^2 B^{(i)}_{B_q} \hspace{0.5in}{\rm and}\hspace{0.5in} \xi = f_{B_s}\sqrt{B^{(1)}_{B_s}}\ / f_{B_d}\sqrt{B^{(1)}_{B_d}},
\label{params}
\end{equation}
where $f_{B_q}$ is the $B$-meson decay constant and ${\mathfrak c}_i$ are numerical factors.\footnote{An alternate definition of the bag parameter includes a factor of $M^2_{B_q}/(m_b+m_q)^2$ with ${\mathfrak c}_i$ for $i \neq 1$.}

\paragraph{\bf The Lattice Calculation}
Working in the $B$-meson rest-frame, we generate correlation function data via Monte Carlo evaluation of the path integral representations of the vacuum expectation values
\begin{equation}
C^{\rm2pt}(t) = \langle B^0_q(t) {B^{0}_q(0)}^\dagger \rangle \hspace{0.4in}{\rm and}\hspace{0.4in} C_i^{\rm 3pt}(t_1,t_2) = \langle B^0_q(t_2) {\cal O}_i(0) B^0_q(t_1) \rangle,
\label{corrfns}
\end{equation}
with gauge field integration performed with the MILC ensembles.  We simulate with staggered light and Fermilab heavy quarks (for details of gluon and quark discretizations see~\cite{Bazavov:2010} and references therein).  Using Bayesian fitting techniques~\cite{Lepage:2002}, these data are fit to the ans\"atze~\cite{Wingate:2003},
\begin{eqnarray}
C^{\rm 2pt}(t) &=& \sum_{n=0}^N Z_n^2 (-1)^{n(t+1)}\left( e^{-M_nt} + e^{-M_n(T-t)} \right) \nonumber \\
C_i^{\rm 3pt}(t_1,t_2) &=& \sum_{n,m=0}^N \langle B^0_q | {\cal O}_i | \bar{B}^0_q \rangle \frac{Z_n Z_m}{2\sqrt{M_nM_m}}(-1)^{n(t_1+1)+m(t_2+1)}\ e^{-M_nt_1-M_mt_2}\ ,
\label{fitfn}
\end{eqnarray}
where $Z_n$ is the amplitude and $M_n$ the mass of the  $B$-meson $n^{th}$ excited state and $T$ the temporal extent of the lattice.  From these fits we extract the matrix elements over a range of valence-quark masses,
light sea-quark masses,
and lattice spacings.
Lattice matrix elements are matched to the continuum at one loop in tadpole-improved lattice perturbation theory~\cite{Gamiz:inprep}.
The continuum matrix elements corresponding to physical light (for $B_d^0$) or strange (for $B_s^0$) valence quark, and at physical light sea-quark mass, are then obtained by extrapolation/interpolation with the aid of rooted, staggered, chiral perturbation theory~\cite{Detmold:2007, Bernard:inprep}.  

\paragraph{\bf Status and Outlook} 
We recently completed a calculation of $\xi=1.268(63)$ using data at lattice spacings down to $\approx$\,0.09 fm and valence quarks as light as $0.1\, m_s$~\cite{Bazavov:2012}. 


%


An ongoing calculation includes an update of $\xi$ and the calculation of matrix elements and bag parameters for all five operators in Eq.~\eqref{mixops}.  In addition, it includes several improvements: a three-fold increase in statistics; data at lattice spacings as small as $\approx$\,0.045 fm and valence quarks as light as $0.05\, m_s$, to reduce the effect of the continuum-chiral extrapolation; and a more thorough treatment of chiral perturbation theory~\cite{Bernard:inprep}.

Preliminary results for the matrix elements with the new, extended data set can be found in Table 4 of Ref.~\cite{Bouchard:2011}. These results are based on data at lattice spacings down to $\approx$\,0.06 fm and valence quarks as light as $0.1\, m_s$. Initial studies show that our new treatment of chiral perturbation theory has a negligible effect on the values of the matrix elements in this analysis.


\begin{theacknowledgments}
Thanks to the organizers for an interesting, well organized conference, to the conveners for the invitation to speak, and to my collaborators:  Elizabeth Freeland, Claude Bernard, Aida El-Khadra, Elvira G\'amiz, Andreas Kronfeld, Jack Laiho, Ruth Van de Water and the rest of the Fermilab Lattice and MILC Collaborations.
\end{theacknowledgments}

\bibliographystyle{aipproc}
\bibliography{087_Bouchard.bib}
\end{document}